\def\selectedoptions{final}

\documentclass[\selectedoptions]{aipproc}

\usepackage{psfig,rotating,amssymb}
\def\selectedlayoutstyle{6x9}
\layoutstyle\selectedlayoutstyle
\SetInternalRegister\hbadness{8000} 

\newcommand\doingARLO[2][]{%
  \ifx\mmref\undefined #1\else #2\fi
}

\newcommand\snts{{SN\,1006}}

\newcommand\arcmin{\mbox{$^\prime$}}%

\newcommand\msun{{$M_{\odot}$}}
\newcommand\net{$n_{\rm e}t$}
\newcommand\kte{$kT_{\rm e}$}
\newcommand\netunit{cm$^{-3}$s}
\newcommand\pcc{cm$^{-3}$}
\newcommand\chandra{{\em Chandra}}
\newcommand\astroe{{\em Astro-E2}}

\newcommand\xmm{{\em XMM-Newton}}

\newcommand\hut{{\em HUT}}
\newcommand\fuse{{\em FUSE}}

\def\nat{{Nat\,}}
\def\adspr{{Adv. of Space Research\,}}
\def\apj{{ApJ\,}}
\def\apjl{{ApJL\,}}
\def\apjs{{ApJS\,}}
\def\aap{{A\&A\,}}
\def\aaps{{A\&AS\,}}

\def\pasj{{PASJ\,}}
\def\aj{{AJ\,}}

\newcommand{\kms}{{km\,s$^{-1}$}}
\begin{document}
\title[High Resolution X-ray Spectroscopy of SN\,1006]
{High Resolution X-ray Spectroscopy of SN\,1006}

\author{Jacco Vink}{
  address={SRON National Institute for Space Research, Sorbonnelaan 2, 3584CA,
           Utrecht, The Netherlands},
  email={j.vink@sron.nl},
}

\copyrightyear  {2004}
\begin{abstract}
I discuss the high resolution \xmm\ Reflection Grating
Spectrometer (RGS)
spectrum of \snts. \snts\ is one of the best examples
of a supernova remnant that is far out of ionization equilibrium.
Moreover, optical, UV and X-ray data indicate that it is also out
of temperature equilibrium.  I discuss the X-ray evidence for this.

In addition I discuss the lower resolution RGS spectrum
of the eastern rim of \snts. Despite the lower resolution, the spectrum
contains significant evidence for an asymmetric expansion velocity.
Two likely solutions fit the O VII triplet. One with no significant
thermal broadening and a shell velocity of $\sim6500$~\kms,
and one with significant broadening and a shell velocity of $9500$~\kms.
The first solution seems the most plausible, as it is consistent with radio
expansion measurements, which suggest a decelerated shell.
\end{abstract}

\date{\today}

\maketitle

\section{Introduction}
The study of supernova remnants lies at the crossroads of astrophysics,
combining the interests in the end products of stellar evolution
with the investigation of the interstellar medium and cosmic ray acceleration.
This is
all the more interesting now that Type Ia supernovae \citep[e.g.][]{tonry03}
have become central to the study of the geometry of the universe and
gamma ray bursts are known to originate from a subset of core
collapse supernovae \citep{stanek03}.

\begin{figure}
\hbox{\psfig{figure=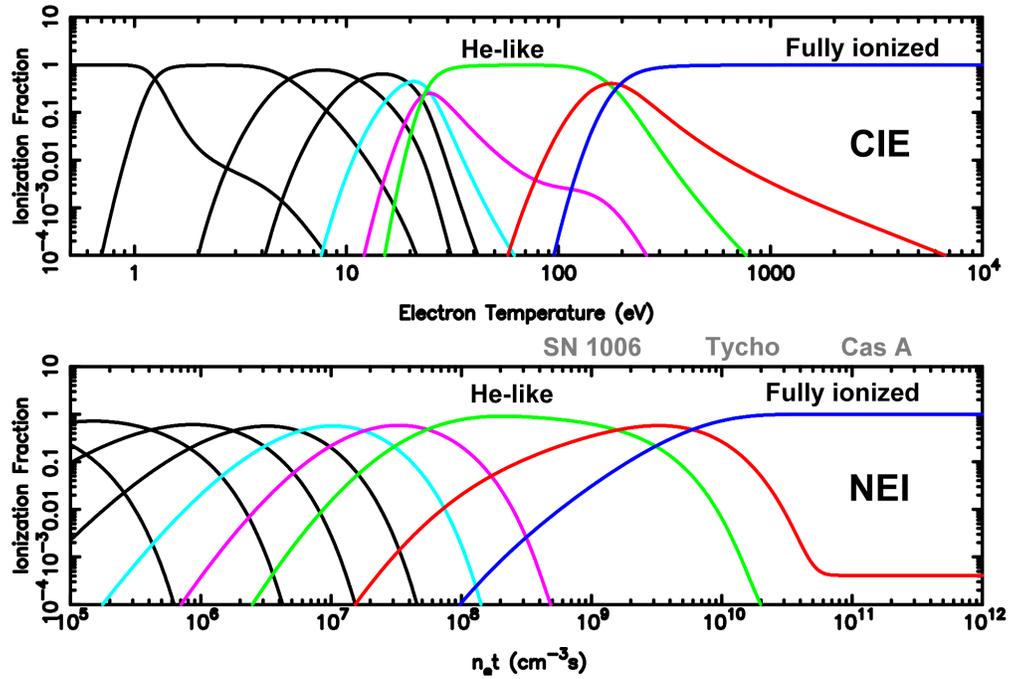,width=0.9\textwidth}}
\caption{An illustration of the effects non-equilibration ionization.
Although the top and bottom panels look very similar, the top panel
shows the oxygen ionization fraction as a function of electron
temperature for the case of collisional ionization equilibrium (CIE),
whereas the bottom panel shows how in the case of recently heated
plasmas (NEI) the ionization fraction at a fixed temperature of \kte$=1.5$~keV
changes as a function of
ionization age, which is the product
of electron density and time since the plasma was heated (\net).
Note the sturdiness of the He-like ion, which is present over
a wide temperature or \net\ range.
The calculations are based on the simple to use, 
but somewhat outdated ionization and recombination rates in \citep{shull82}.
}
\label{ionization}
\end{figure}

\begin{figure}
\psfig{figure=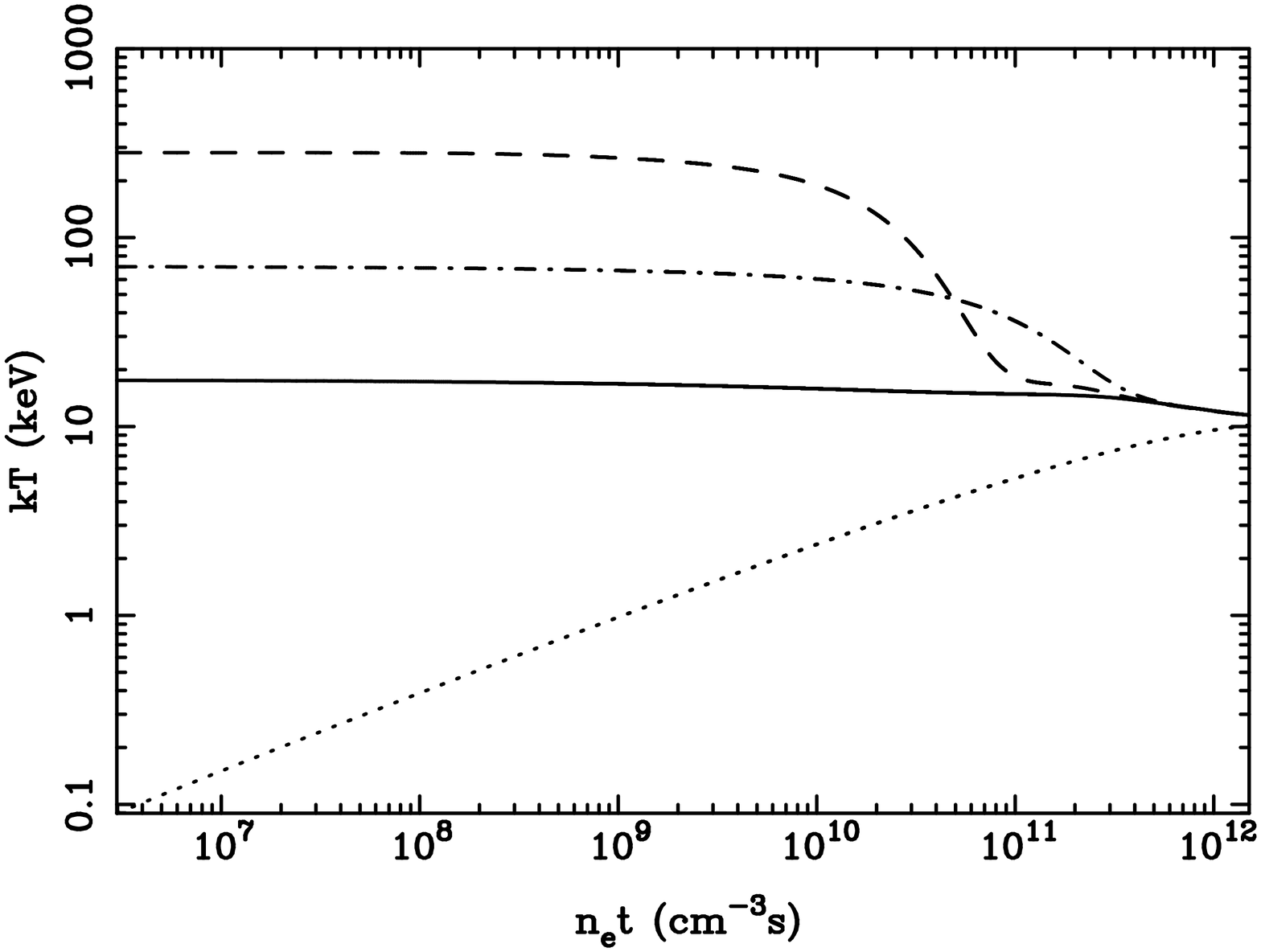,angle=-90,width=0.6\textwidth}
\caption{An illustration of the effect of non-equilibration of temperatures, 
taking into account ionization effects, but not adiabatic expansion.
The temperature of the electrons (dotted), protons (solid), helium 
(dashed-dotted) and oxygen ions (dashed) is proportional to their mass ratio 
behind the shock.
Due to Coulomb interactions the temperatures slowly equilibrate. 
It is interesting to note that the oxygen-proton equilibration will 
eventually be faster than the helium-proton equilibration, 
as the oxygen ions acquires a higher charge and the interactions
scale with $Z^2$\ 
\citep{zeldovich66,nrlplasma}.
}
\end{figure}

However, of more interest for these proceedings
is that supernova remnants are objects in which interesting plasma processes
take place.
Supernova remnant shocks move through a medium with a
typical density of 1~\pcc. At those low densities Coulomb (particle-particle)
interactions are rare, leading to typical collision time scales
of several thousand years. 
Shock heating does, however, occur in supernova remnants, 
otherwise we would not observe X-ray emission from hot plasmas. 
This implies that the heating process takes place through collective effects, involving
plasma waves.
Apart from being collisionless, supernova remnant shocks are also
the most probable source of cosmic rays,
at least for energies up to $\sim 10^{15}$~eV, but possibly even up to
$\sim 10^{18}$~eV \citep[][]{bykov04,vink04b}.

The fact that supernova remnant shocks are collisionless and are
the locations of particle acceleration has two important consequences.
First of all, we can no longer assume that different particles species
(protons and other ions and electrons) are in temperature equilibration.
In an extreme case, and ignoring particle acceleration, 
the temperature, $kT_i$, of each plasma component, $i$, is:
 
\begin{equation}
        kT_{i} = \frac{2(\gamma-1)}{(\gamma+1)^2} m_i v_s^2  = 
	\frac{3}{16} m_i v_s^2,
       \label{eq-shocks}
\end{equation}
where $\gamma$ is the adiabatic index, 
$m_i$, is the particle mass, and $v_s$ is the shock velocity.
For full equilibration this is $kT = 3/16 <m>v_s^2$.
In addition,
we have to consider that efficient
cosmic ray acceleration, and possible magnetic field generation, will take
away energy, lowering the plasma temperature(s) of the thermal plasma 
components \citep{hughes00b}.

As a result measuring the plasma temperature through X-ray spectroscopy,
which usually will give us the {\em electron} temperature,
will no longer inform us directly about the shock velocity.
This has been known for quite some time \citep[e.g.][]{itoh77,itoh84},
but until recently it was ignored consciously \citep[e.g.][]{jansen88} 
or subconsciously, as it was difficult to assess the amount of 
temperature equilibration from the observational data.

However, another form of non-equilibration, 
namely non-equilibration of ionization (NEI), has received more attention
over the last three decades, because its consequences could be more easily
deduced from X-ray spectra of sufficient quality 
\citep[e.g][]{winkler81,
gronenschild82,hughes85,jansen88}.

The concept of  NEI is relatively simple \citep{itoh77,mewe80,liedahl99}.
NEI is important in supernova remnants as
in the relatively recently heated plasmas the ions have not had 
sufficient time to ionize up to the ionization 
level at which the number of ionizations is compensated 
by the number of recombinations 
(i.e. collisional ionization equilibration or CIE). 
The effect of NEI is illustrated in 
Fig.~\ref{ionization}. Observationally NEI gives rise to a mismatch between
the electron temperature derived from line emission ratios,
and the electron temperature derived from the continuum shape, which reflects
the actual electron temperature.
In addition, spectra of NEI plasmas will display lines that are unique
for NEI, and are the result of 
inner shell excitations and ionizations.
The Fe-K line emission is most widely used as a diagnostic tool
for NEI conditions, as for a plasma with \kte$ \gtrsim 2$~keV the centroid
of the $n=2\rightarrow1$\ line emission shifts gradually
from $\sim6.4$~keV to $\sim 6.7$~keV for Fe XXV \citep{palmeri03,mendoza04}.

Here I discuss high resolution spectra of a supernova
remnant that is one of the most extreme cases of non-equilibration
of ionization and temperatures, \snts.
 
\begin{figure}
\hbox{
\vbox{
\psfig{figure=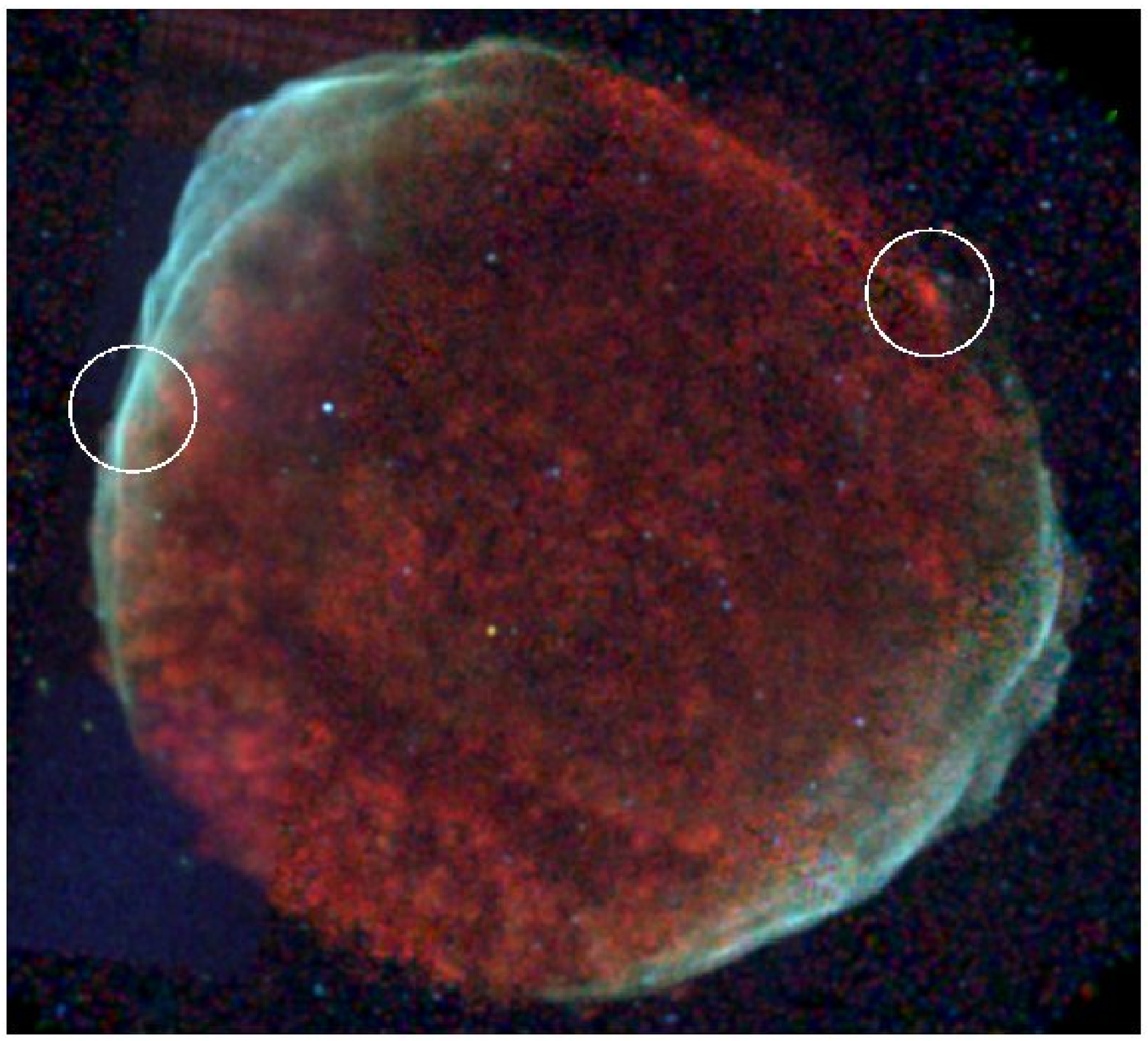,width=0.5\textwidth}
\vskip 4mm
}

\psfig{figure=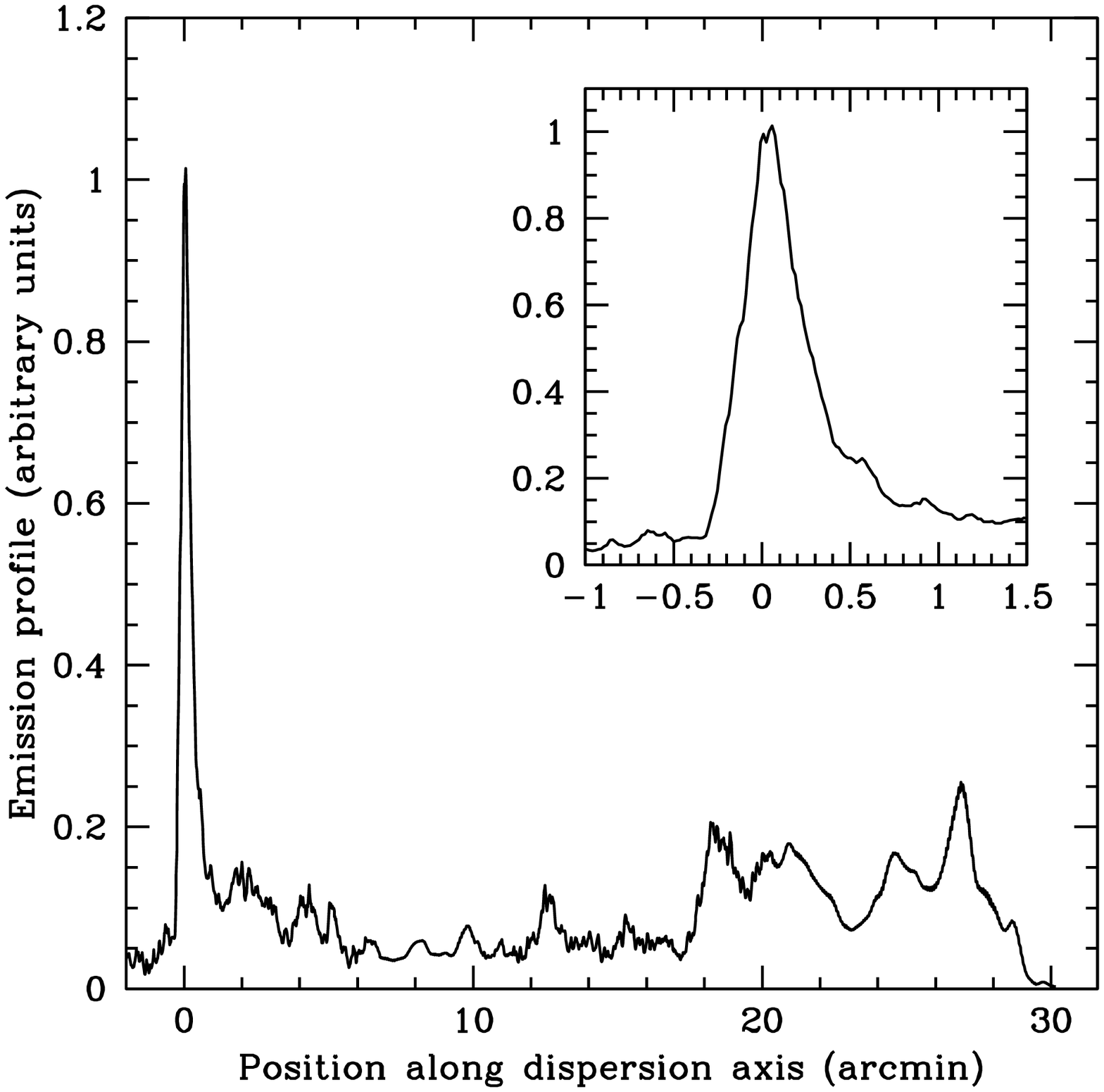,width=0.5\textwidth}
}
\caption{\chandra\ mosaic of SN 1006. The circles indicate the \xmm\ pointings discussed here.
The emissivity profile is taking in the dispersion direction of the 
RGS
\citep[from][]{vink03b}.}
\label{image}
\end{figure}

\begin{sidewaysfigure}
\hbox{
\psfig{figure=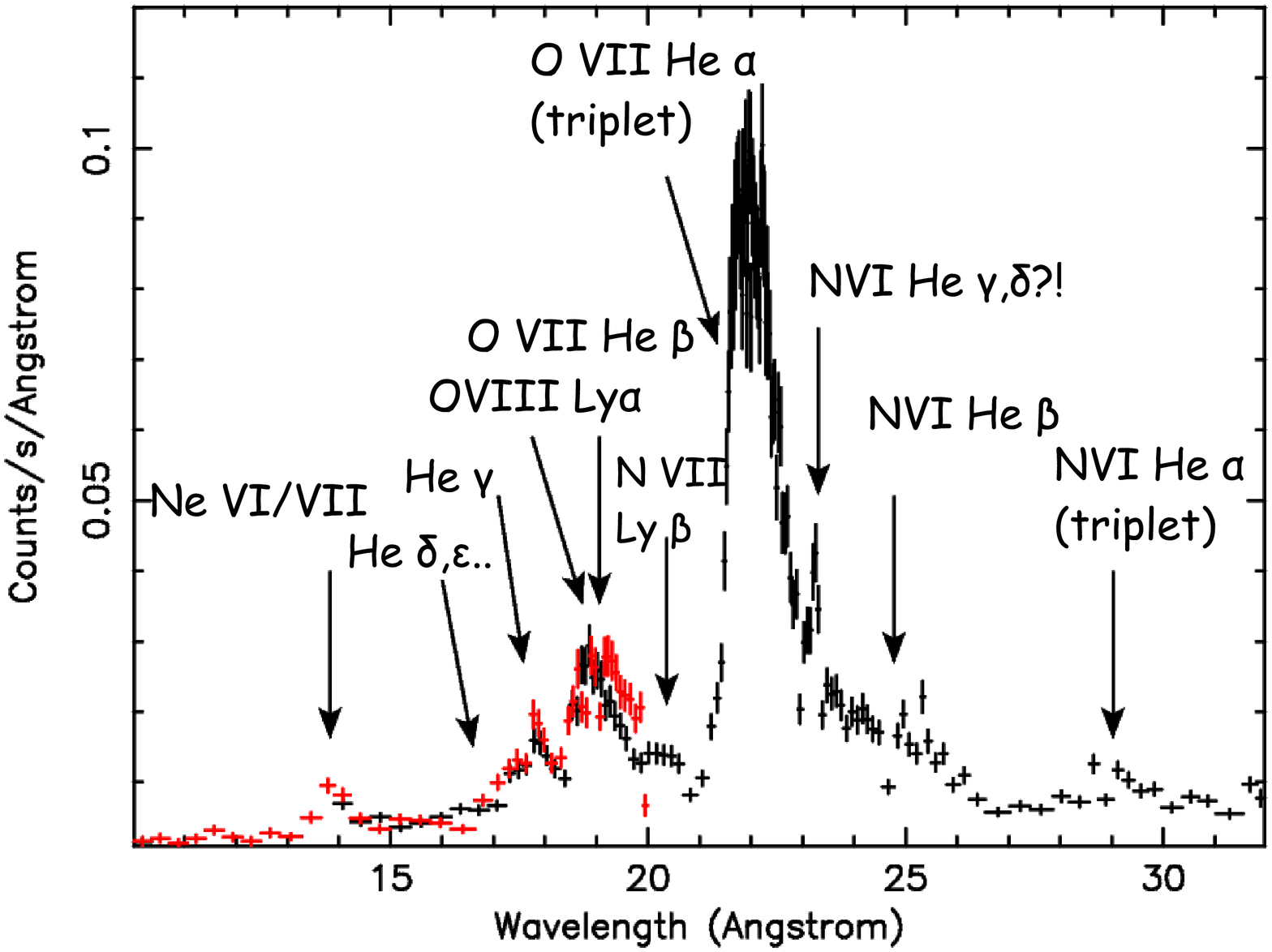,width=0.9\textheight}
}
\caption{The \xmm-RGS spectrum of the northwestern knot.}
\end{sidewaysfigure}

\section{The \xmm-RGS spectra of SN 1006}
The young historical remnant SN 1006 is thought to be a Type Ia supernova
remnant on account of its high Galactic latitude (+14.6)
and the long duration of the historical light curve \citep{stephenson02}.
The high latitude also explains the relatively low density of the local 
medium of
$\sim 0.1$~cm$^{-3}$\ 
\citep[][]{vink00a,allen01,dyer04,long03}.
This makes \snts\ the ideal candidate for studying the two non-equilibration 
effects,
as it has an ionization parameter of \net$\sim 2\times 10^9$~\netunit. 
The distance to \snts\ is 2.1~kpc \citep{winkler03}.

\xmm\ observed \snts\ several times, both as a guaranteed time and guest 
observer target. 
I will discuss here the guest observer observations of August 2002, which consisted
of two pointings. 
One pointing was on the bright knot at the northwestern edge of the remnant,
the other was on a narrow feature on the eastern rim (Fig.~\ref{image}).
SN 1006 is a large remnant (30\arcmin), but the pointings were optimized
in order to obtain high resolution spectra of bright, small structures.
Moreover, the pointings were on the rest of the remnant where it is unlikely that
there are bulk plasma motions in our line of sight. This is necessary in order to
allow estimates of the thermal line broadening.

A sufficiently high resolution spectrum was obtained for the
northwestern pointing, but for the eastern rim, only the non-thermal X-ray emission is very narrow, whereas
the thermal emission is much more extended, degrading the spectral resolution.
This is somewhat unfortunate, as the electron and ion temperatures of the plasma may be
influenced by efficient cosmic ray acceleration, and non-thermal emission from the rims of SN 1006
indicate a very efficient electron cosmic ray acceleration at the west and eastern rims, but
not at the northwestern edge \citep{koyama95,rothenflug04}.

\begin{figure}
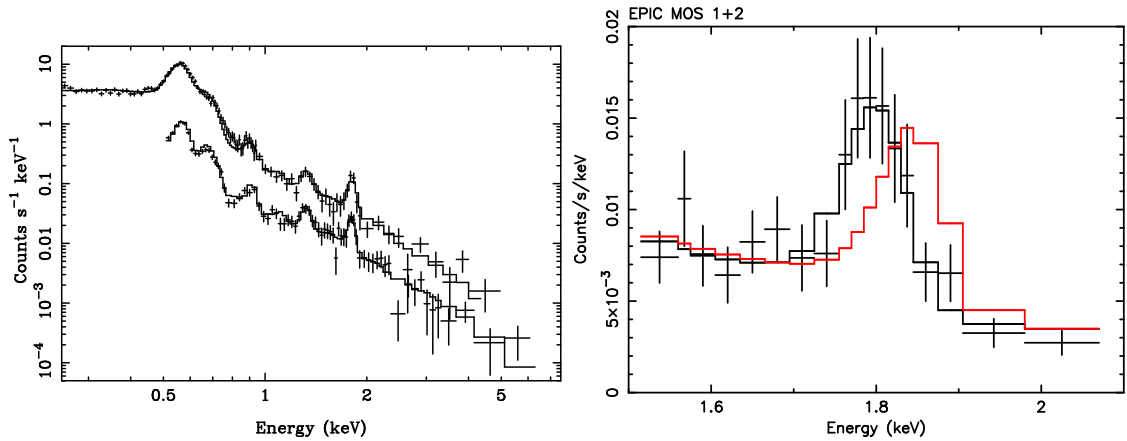

\hbox{
  \psfig{figure=sn1006nw_epic_brspot.ps,angle=-90,width=0.5\textwidth}
  \psfig{figure=sn1006_nwknot_mos12_SiK.ps,angle=-90,width=0.5\textwidth}
}
\caption{The \xmm\ PN and MOS spectra (left) and a detail of the Si emission as observed
by the MOS detectors (right). The shifted line shows the position of the Si 
emission if most of the Si would have been in the He-like state. }
\label{ccdspectra}
\end{figure}

\subsection{The X-ray spectrum of the northwestern knot}

The \xmm\ Reflection Grating Spectrometer \citep[RGS,][]{denherder01} 
spectrum of the northwestern knot shows
many spectral details that cannot be seen in CCD spectra. 
The most dominant emission is
coming from O VII, in particular from the O VII triplet, but also O VII line 
emission
from higher quantum levels up to the series limits. 
The line emission around 19~\AA\ is a mix of O VII He$\beta$\ emission and 
O VIII Ly$\alpha$
emission. Lines from Fe-L transitions, in particular Fe XVII which has lines
around 15~\AA and 17~\AA\ are conspicuously absent.
This may be surprising given the fact that SN\ 1006 is a likely Type Ia 
remnant, which produce $\sim 0.7$~\msun\ of Fe. However, for the low ionization
age of \snts\ Fe is expected to have a peak ionization state of Fe XIII,
so no Fe XVII emission is to be expected.

The CCD spectra were modeled using the SPEX\footnote{
\url{http://www.sron.nl/divisions/hea/spex/version2.0/index.html}} 
\citep{kaastra96} spectral code. 
The detailed results can
be found in \citep{vink03b} and confirm the low ionization age of the plasma, 
\net $= 2.4\times10^9$~\netunit, whereas \kte$=1.5\pm0.2$keV.
There is a relative overabundance of Si, which suggests
that the knot is an ejecta knot, and not a shocked pre-existing density 
enhancement
in the interstellar medium. Interestingly, there is a weak enhancement of 
X-ray emission
in front of the knot as well, that seems to emit slightly harder X-ray emission.
It is not clear how these relate to each other.

\begin{figure}
\hbox{
 \psfig{figure=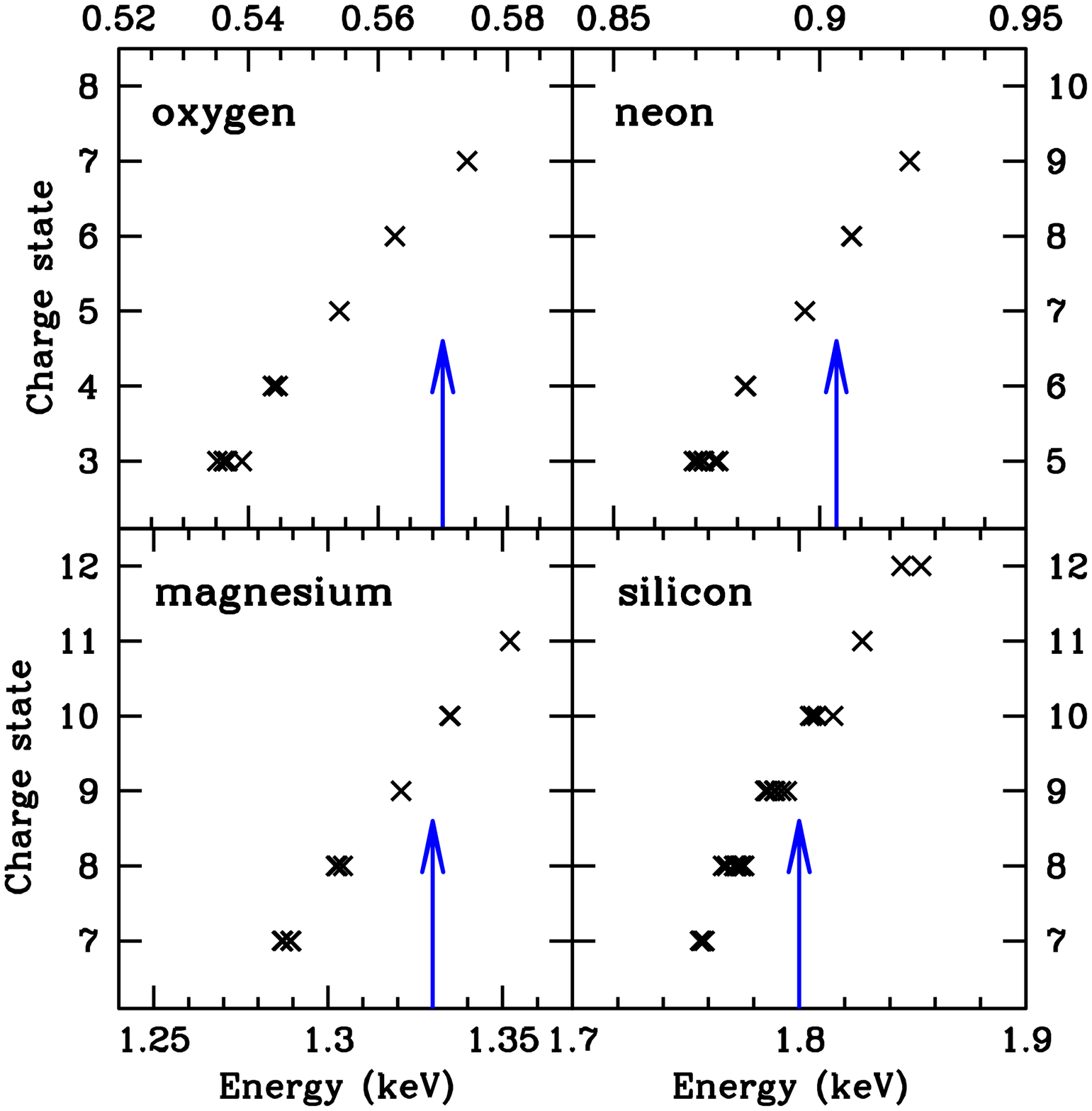,width=0.5\textwidth}
 \psfig{figure=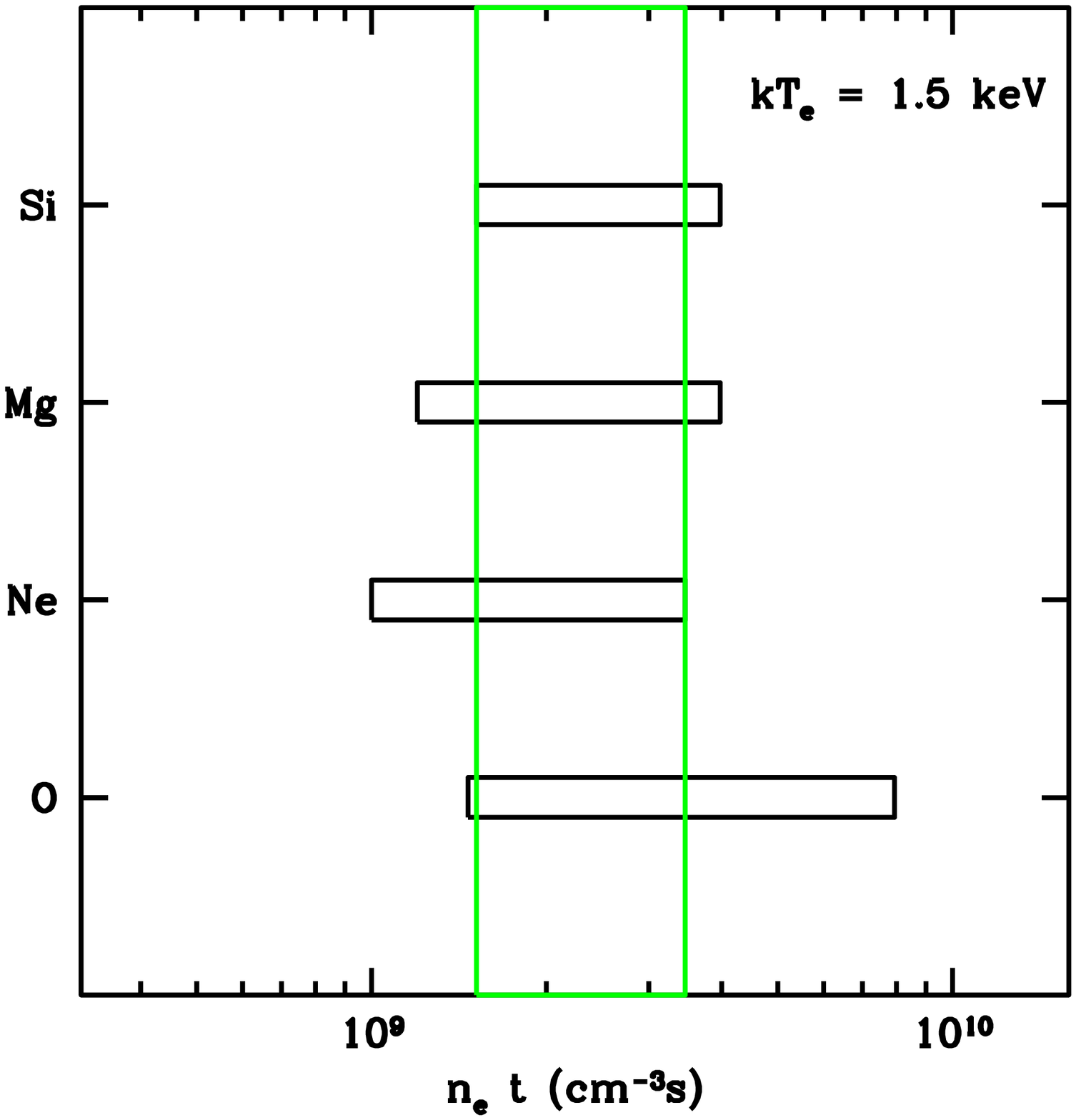,width=0.5\textwidth}
}
\caption{
Line
energy versus charge for O, Ne, Mg, and Si. The crosses give the expected
line energies for different ionization states (from Dr. M.F. Gu's FAC code) 
and the
arrows approximately indicate the observed centroid energies.}
\label{centroids}
\end{figure}

The low ionization age of the plasma can be deduced not only by globally 
fitting the spectrum, but
also by measuring the centroid of each individual line, or by modeling the 
O~VII triplet. 
For instance the EPIC PN and MOS spectra clearly indicate that the centroid of 
the Si emission
is $1.80\pm0.01$~keV instead of at 1.85~keV, which is the centroid expected if 
Si is in the He-like
state (Fig.~\ref{ccdspectra}). Also the Ne and Mg line centroids are at lower 
energies
than the He-like.  Purely combining the information of the centroids, 
and assuming
\kte $=1.5$~keV one also finds an ionization age of 
\net\ $\sim 2\times10^9$~\netunit\ 
(Fig.~\ref{centroids}). 
The most important inner shell lines are present in the SPEX spectral code,
but in general one should be aware that spectral codes are not always
complete in this sense, as most atomic databases were initially compiled with
CIE plasmas in mind. Moreover, the line parameters may also be less accurate,
as less experimental data exists to measure the line properties.

\subsection{Non-equilibration of temperatures in the northwestern knot}
One of the main goals of the \xmm\ observations was to measure the amount of 
non-equilibration
of electron and ion temperatures, by measuring the thermal Doppler broadening 
of O VII and O VIII
line emission. So far evidence for non-equilibration in supernova remnants 
comes mainly from measuring
the H$\alpha$\ line width and broad to narrow line ratios.
This has been done for several remnants including \snts\ 
\citep{ghavamian02,rakowski03}. 
The amount of non-equilibration
has also been measured in the UV from C IV, N V and O VI lines by 
\hut\ \citep{laming96} and very recently by \fuse\ \citep{korreck04}. 
The optical and UV results indicate that there is no equilibration at the 
shock front, although  the \fuse\ results
suggests that the temperature of different species is not proportional to the 
species mass, as implied by eq.~\ref{eq-shocks} and the \hut\ spectrum.

\begin{figure}
\psfig{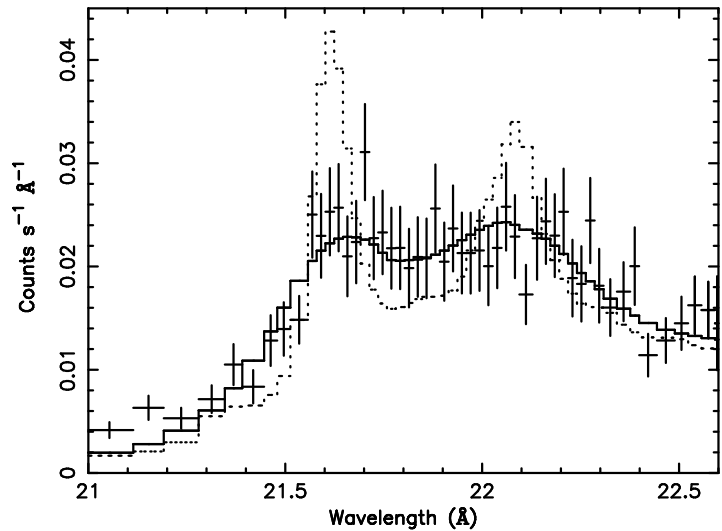}
\caption{The O\,VII triplet from the northwestern not as observed by the RGS1. 
The dotted line shows the best fit spectrum with (solid) and without including 
thermal Doppler broadening.}
\label{broadening}
\end{figure}

The O VII He$\alpha$\ emission consists of the He triplet, plus a potential 
contribution from an O VI line very close
to the O VII forbidden line emission.
In order to measure the Doppler broadening models for the line ratios
for a grid of \net\ values were fitted, using parameters obtained from
the flexible  atomic code (FAC) of Dr. M. F. Gu \citep{gu03}.
The spectrum of the RGS1 O VII spectrum and the best fit models are shown in 
Fig.~\ref{broadening}.
The best fit suggests a very significant broadening at a confidence level of  
9.8$\sigma$.
Allowing full freedom of the line normalizations still indicates thermal 
broadening, at a still high confidence level
of 6.5$\sigma$. The broadening itself is $\sigma_E = 3.4\pm0.4$~eV. 
The published results are based on a response matrix that was corrected for 
the emissivity  profile shown in Fig.~\ref{image}. 
However, the fitting program XSPEC also contains the convolution model 
{\em rgsxsrc} made by Dr. A. Rasmussen, which uses an input image to determine
the emissivity profile. Using {\em rgsxsrc} gives very similar results.

The derived value for the broadening implies an oxygen temperature of 
$kT = 530\pm150$~keV, which looks incredibly high,
but is in effect consistent with eq.~\ref{eq-shocks} if the gas was 
shocked with a velocity of $\sim 5000$~\kms.
Of course the oxygen temperature is more than a factor 300 higher than the 
electron temperature, clearly indicating non-equilibration of temperatures.
The measured O VII line broadening is more than reported for O VI UV emission 
from other regions of the northwestern rim 
\footnote{Unfortunately in \citep{korreck04} the measured broadening in 
$\sigma$ is mistakenly taken to be a FWHM.} \citep{korreck04}.
There are several reasons why there could be a discrepancy. 
The knot may be an ejecta knot, as implied by the high Si abundance, 
and the heating may have been caused by the reverse shock, 
which has a different speed than the shock.
Another possibility is that the plasma of the knot was shocked earlier than 
the regions observed in O VII, after all the X-ray
measurements concern more highly ionized material, which is 
situated further downstream from the shock.

\subsection{What about the eastern rim?}

The eastern and western rims of SN 1006 are likely sites of efficient cosmic 
ray shock acceleration \citep{koyama95}.
If the cosmic ray pressure dominates over the thermal pressure, 
eq.~\ref{eq-shocks} is no longer 
valid. It will result in an average plasma temperature that is lower. Moreover,
if the cosmic ray is very efficient and the cosmic ray power law
index is equal to or flatter than -2, the energetic cosmic rays that escape 
the shock will drain the shock of its energy and the shock will become 
(cosmic) radiative, resulting in much higher shock compression ratios
\citep{ellison04}.

Given the interesting physics at the eastern and western rim, it is 
unfortunate that the RGS spectrum from the eastern rim
does have a high spectral resolution,
especially since no UV emission lines from the eastern rim have been 
detected by \fuse\ \citep{korreck04}.
It is therefore still worthwhile to obtain as much information from 
the RGS spectrum as possible. I will therefore discuss here a preliminary 
analysis of RGS spectrum.

The effective spectral resolution is mainly determined by the width of the 
shell, about 5\arcmin, resulting in an effective resolution for the first 
order spectrum of $\sim0.6$~\AA.
This is not sufficient to resolve the O VII triplet, which is also for the 
eastern rim the dominant line emission. However, there is still high spectral 
resolution information in the spectrum, which results from the sharp increase 
of the emission at the edge of the remnant.

\begin{figure}
\vbox{
\hbox{
\vbox{
 \psfig{figure=sn1006e1_0kms.ps,angle=-90,width=0.5\textwidth}
\vskip 2mm
}
 \psfig{figure=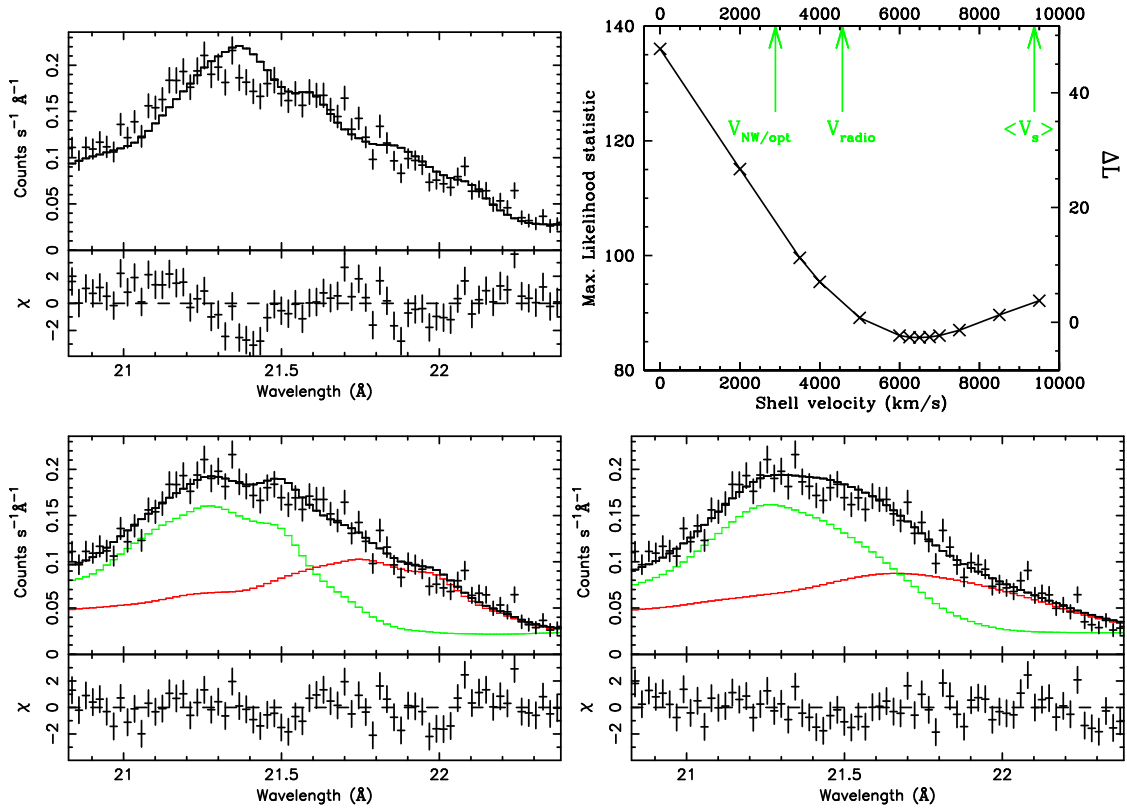,angle=-90,width=0.5\textwidth}
}
\vskip 2mm
\hbox{
  \psfig{figure=sn1006e_6500_f0.7_combined.ps,angle=-90,width=0.5\textwidth}
  \psfig{figure=sn1006e_9500_f0.7_combined.ps,angle=-90,width=0.5\textwidth}
}}
\caption{
The O VII triplet from the eastern rim as observed by RGS1.
Top left: The best fit model with only taking into account
the spatial profile.
Top right: the likelihood statistic as a function of shell velocity,
for a front back asymmetry of 70\%. The arrows indicate other velocity
measurements from respectively the northwest, radio and the average velocity.
Bottom: The left panel shows the best fit solution
for a shell velocity of 6500~\kms with no broadening. The right panel
shows a solution with a velocity of 9500~\kms\ including substantial
thermal broadening. }
\label{eastern}
\end{figure}

The O VII triplet emission is displayed in Fig.~\ref{eastern}.
It shows that a model that does only contain the triplet lines
and an instrumental response that only takes into account the spatial
emission profile does not give an adequate fit.
The long wavelength side seems to fit reasonably well,
but there is a dip in the data around 21.5~\AA, and the model underpredicts 
the emission around 21.3~\AA. Note that here the interior of remnant is 
dispersed in an opposite direction compared to
the spectrum of the northwestern knot. So the edge of the remnant determines
the shape of the long wavelength side. This suggests that discrepancy is 
caused not so much by line broadening, but by line shifts. 
The plasma motions at the  edge of the remnant, will be mostly in the plane of 
the sky, whereas going to the interior of the remnant radial
motions will give rise to Doppler shifts.
As we are observing both the front and backside of the shell, 
both blueshifts and redshifts should be taken into account.
The redshifts are not directly noticeable 
(they blend in with the rest of the triplet emission), 
but the blueshifts can be seen as the mismatch between model and spectrum 
around 22.3~\AA.

I have attempted to fit a simple model for the O VII triplet, 
taking into account the 
increase in blueshifts and redshifts moving interior to the remnant
(corresponding to shorter wavelength). 
As it is a mixing of spectral and spatial information, 
the method consisted in making a grid of response
matrices in which the spatial and redshift and blueshift were incorporated.
The input to the model consisted of the projected plasma velocity, 
which increased toward the interior, 
and a parameter that determined how much of the emission was coming from the
near and far side of the remnant.
This is a simplification, as velocity gradients are presents 
and the morphology of
the shell may be complicated. It was nevertheless possible to obtain 
acceptable fits to
the O VII triplet emission (Fig.~\ref{eastern}).

It turns out there are two good fitting models.
One with no, or only a small additional (thermal) broadening 
($\sigma_E = 0.4\pm0.3$~eV), and
a shell with
velocity $6500\pm830$~\kms\ (68\% confidence range, Fig.~\ref{eastern}).
The other solution is for a much higher shell velocity of $> 9000$~\kms,
and quite a substantial additional broadening of $\sigma_E = 3\pm1$~eV.
In both cases the best solution suggests that about 70\% emission
comes from material moving toward us ($>55\%$ at a 3$\sigma$ confidence).
The solution with a large broadening gives a better fit (Fig.~\ref{eastern}),
but is less plausible. The reason is that a bulk velocity of $>9000$~\kms
would imply that the shock heated plasma is still in free expansion,
whereas a solution with $v\sim6500$~\kms\ fits in much better 
with radio proper motions \cite{moffet93}.
The average shock velocity of \snts, given its age and distance is 
$<v>=9375$~\kms. 
The radio expansion measurements suggest that 
$v_s = (0.48\pm0.13)<v> = (4500\pm1220)$~\kms. 
Both solutions exclude the possibility that the plasma velocity is
similar to what has been measured for the northeastern H$\alpha$\
filaments $v_s = 2800$~\kms\ \citep{ghavamian02}. This is consistent with
the smaller radius in the northwest and the finding of an increased
density as indicated by
21cm line measurements \citep{dubner02}.

The shape of the O VII triplet suggests that more material
is blueshifted than redshifted. This asymmetry is
reminiscent of the difference in forward shock and reverse shock
velocity found in absorption toward a  UV star projected on the center
of \snts\ \citep{hamilton97}. The reverse shock appears to have penetrated
more into the innerejecta layers of the front side than on the backside.
Such an asymmetry is likely to 
be the result of a higher density at the front side,
which will also result in more shocked heated plasma at the front side.

\section{Concluding remarks}

\snts\ is from many points of view an interesting supernova remnants.
It is one of the few historical remnants, and it was the first remnant
for which it was established that the particle acceleration by the shock
was efficient enough to produce X-ray synchrotron emission.
The fact that \snts\ expands into a low density medium makes it
also an ideal laboratory to study the effects of 
non-equilibrium ionization and non-equilibration of temperatures,
both of which depend on the value of \net, which is around 
$2\times10^9$~\netunit\ for \snts, one of the lowest values found
in a supernova remnant.

The measurements presented here show the diagnostic power of high resolution
spectroscopy when it comes to characterizing the ionization
state, temperature and velocities of hot plasma. 
The results on a remnant like \snts\ 
raises the expectation for the results from micro-calorimeters carried
by \astroe, and {\it Constellation-X/Xeus},
which will make it much more straightforward to measure Doppler motions
from extended sources.

I thank the organizers of the XDAP 2004 workshop for their invitation and
financial support.

\end{document}